# Ultrafast Kilowatt-Range Microwave Pulsing for Enhanced $CO_2$ Conversion in Atmospheric-Pressure Plasmas

S. Soldatov[1], L. Silberer[1], C.K. Kiefer[3,4,5], G. Link[1], A. Navarrete[2] and J. Jelonnek[1]

[1] Institute for Pulsed Power and Microwave Technology (IHM), Karlsruhe Institute of Technology (KIT), Kaiserstraße 12, Karlsruhe, 76131, Germany
[2] Institute for Micro Process Engineering (IMVT), Karlsruhe Institute of Technology (KIT), Kaiserstraße 12, Karlsruhe, 76131, Germany
[3] Work done while empolyed at Max-Planck-Institut für Plasmaphysik, 85748, Boltzmannstraße 2, Garching, Germany
[4] Current affiliation: Research group PLASMANT and Center of Excellence PLASMA, Department of Chemistry, University of Antwerp, Universiteitsplein 1, Antwerp, 2610, Belgium
[5] Current affiliation: Electrification Institute, University of Antwerp, Olieweg 97, 2020 Antwerp, Belgium

**Abstract.**

Ultrafast microwave power pulsation is demonstrated as an effective strategy to enhance $CO_2$ conversion in atmospheric-pressure plasma reactors. While initial experiments at several hundred watts in a compact coaxial plasma torch showed improved performance, the present study investigates the scalability of this approach to kilowatt-range microwave power. Conversion and energy efficiency were examined in two reactor configurations: a Surfaguide-based system (KIT) and a cavity-based plasma torch (IPP), and benchmarked against the compact coaxial torch. Both kilowatt-scale setups share similar microwave coupling schemes, power levels, reactor tubes, and gas injection geometries, but differ in afterglow treatment. The torch at IPP employs rapid nozzle-based quenching, whereas the Surfaguide-based reactor relies on slower cooling along an extended quartz tube. Stable plasma operation was achieved at pulsation peak powers of ~4 kW and pulse durations from sub-microseconds to microseconds, with stability limited to inter-pulse times of ~10 μs (cavity-based torch) and ~12 μs (Surfaguide-based reactor). In contrast to the coaxial torch, no plasma reignition regime was observed in either kilowatt-scale reactor, resulting in weaker plasma temperature modulation. Notably, the period-averaged gas temperature in the Surfaguide-based reactor exceeded that under continuous-wave operation. Under these conditions, relative enhancements of <40% in $CO_2$ conversion and <20% in energy efficiency were measured compared with continuous-wave operation. These improvements were largely suppressed in the torch at IPP, presumably due to rapid afterglow quenching. Finally, analysis of the instantaneous reflected microwave power provided qualitative insights into electron density dynamics during the power-OFF and power-ON phases.

## 1. Introduction

The capture of carbon dioxide ($CO_2$) from the atmosphere or from industrial exhausts with its consecutive conversion into value-added chemicals is believed to reduce greenhouse gas emissions on our planet and mitigate global warming. Particularly, the electro-chemical conversion of $CO_2$ when it is driven by electricity from renewable energy sources, like wind and solar, draws special attention. The use of renewable electricity, which is intermittent by nature, for production of synthetic hydrocarbons from captured $CO_2$ allows the storage of surplus electrical energy from renewables when the supply exceeds the demand. Liquid fuels synthesized with captured $CO_2$ and renewable electricity can replace the fossil-based kerosene [1], gasoline and diesel and enable the closing of the carbon cycle in the transport sector. The two main electro-chemical approaches for $CO_2$ conversion are based on electrolysis and plasma. Plasma-assisted $CO_2$ conversion features a quick start and stop of process that matches fluctuating renewable electricity very well [2], high energy density, and no use of rare-earth materials. Among different types of plasma-based reactors, microwave (MW) plasma reactors have shown the most promising results so far with energy efficiency of up to 90% [3-5], though at vacuum conditions. The operation of microwave plasma reactors at atmospheric pressure instead of vacuum is favorable for the industrial applications, because it reduces the energy consumption and capital expenditures compared to vacuum systems. However, the $CO_2$ conversion decreases with increasing pressure. For the $CO_2$ splitting reaction which reads as

$$CO_2 \rightarrow CO + \frac{1}{2}O_2 + 2.93\ eV, \tag{R1}$$

the deterioration both conversion and energy efficiency in microwave plasma discharge at pressures above 0.3 bar is mostly linked to the CO recombination reactions [6]:

$$CO + O + M \rightarrow CO_2 + M \tag{R2}$$

$$CO + O_2 \rightarrow CO_2 + O, \tag{R3}$$

where $M$ stands for a collision partner. Berthelot *et al* has shown that along with increasing pressure and temperature the rates of reactions R2 and R3 increase and $CO_2$ conversion degrades. For microwave reactors operating at near atmospheric pressure the extremely high gas temperatures (6000 -7000 K) [7-10] are unavoidable while $CO_2$ plasma at p>0.2÷0.3 bar experiences the transition from diffuse to contracted regime, and its temperature rises steeply [7]. This phenomenon of plasma contraction described in [11] is characterized by formation of a hot plasma filament whose cross-section size can be significantly smaller than diameter of reactor tube [7, 12]. At such plasma conditions, the rapid quenching is crucial to avoid the deterioration in conversion and efficiency [6, 13]. Particularly in microwave plasmas, experiments employing active gas quenching via water-cooled metallic structures in the afterglow, such as a "gas shower" [14] and nozzle [15-17] have been reported, demonstrating a significant increase in reactor performance compared with setups without quenching.

Alternatively, to increase conversion and process efficiency at near-atmospheric pressures, the microwave (MW) power sustaining the plasma can be delivered in short pulses rather than continuously. Microwave pulsation has been used in different reactor configurations for investigation both plasma dynamics [18-24] and chemical conversion [9, 25-29]. Because of instant microwave-to-electrons energy coupling, the electron density, $n_e$, and temperature, $T_e$, are found to respond promptly to microwave power modulation [20, 25, 30]. The response in $T_e$ was found to be much faster than in $n_e$ and, if the inter-pulse time was long enough, the overshoot in electron temperature in the beginning of pulse was documented [20, 30], Such $T_e$ and $n_e$ dynamics were also confirmed in global plasma model with pulsed energy [31] and multi-physics modelling for hydrogen plasma and pulsed microwave energy in [22]. Additionally, in experiments with $H_2$ plasmas (Rousseau et al 1994 [18]) and in global model of Lieberman and Ashida for argon and chlorine plasmas [31] a higher mean electron density was reported in pulsed discharge as compared with continuous discharge. Moreover, the modulation of vibrational and gas temperatures in pulsed plasma experiments [21, 22, 25] and their relation to the energy efficiency supported with plasma modelling [32-34] were reported.

Vermeiren and Bogaerts in 0D model demonstrated that vibrationally driven dissociation in $CO_2$ plasma can be optimized by power pulsing through appropriate adjustment of the pulse duration and inter-pulse time at a pressure of 0.1 bar, a reduced electric field of 50–150 Td, and a specific energy input (*SEI*) of approximately 1 eV molecule$^{-1}$[33]. In that work, optimal pulse and inter-pulse times were found to result from a trade-off between the time required to reach maximum overpopulation of high vibrational levels and the vibrational–translational (V–T) relaxation time, which decreases with increasing gas temperature. Experimentally, a similar optimum was identified for an atmospheric MW plasma in a coaxial plasma torch at $0.02 < SEI < 0.22$ eV molecule$^{-1}$ [9]. In that study, power pulsation enabled a substantial enhancement in conversion and efficiency compared with continuous plasma operation, exceeding 100%. In addition, clear evidence of non-equilibrium between the rotational ($T_{rot}$) and vibrational ($T_{vib}$) temperatures was observed, with a maximum ratio of $T_{vib} / T_{rot} \approx 2$. These findings motivated pulsation experiments in larger plasma reactors operating at kilowatt MW power and at *SEI* values from 1 to 5 eV molecule$^{-1}$, using two reactor configurations: (i) a Surfaguide-based [35] reactor installed at Karlsruhe Institute of Technology (KIT), and (ii) a modular plasma torch consisting of a cylindrical and coaxial resonator allowing plasma self-ignition installed at Max Planck Institute for Plasma Physics (IPP) [15, 36, 37] which is referred further in text as cavity-based plasma torch at IPP.

In the present paper, we perform a detailed comparison of experimental results obtained in the two abovementioned configurations and relate them to results from pulse-plasma coaxial torch experiments reported in [9, 26]. We investigate how kilowatt-range MW power pulsations with durations ranging from several hundred nanoseconds to several microseconds influence $CO_2$ conversion and energy efficiency in atmospheric microwave-sustained plasmas. The rest of paper is organized as follows. The experimental setups and diagnostic methods used in the three abovementioned studies are described in Section 2. The results are compared and discussed in Section 3, and conclusions together with an outlook for future experiments are presented in Section 4.

## 2. Setups and experimental methods

This section describes three microwave-sustained $CO_2$ plasma setups, all employing solid-state microwave (SSM) generators capable of operating in continuous-wave (CW) and power-modulated modes with pulse and inter-pulse durations of ≥ 50 ns. In the coaxial torch experiment a 300 W (maximum rated power) SSM generator with pulse rise and fall times of ~30 ns was used [9], whereas the in the Surfaguide-based and IPP experiments we employed a 4 kW (maximum rated power) SSM generator with pulse edge times of ~15 ns [38]. In both cases, the microwave power is adjustable in discrete steps over the ranges 20–300 W and 20–4000 W, respectively, and the operating frequency is tunable stepwise within 2.4–2.5 GHz. Frequency tuning is essential for reliable plasma ignition in the coaxial torch configuration.

*2.1 Coaxial microwave torch experiment (KIT)*

Details of the coaxial plasma torch experiment have been reported previously in [9]. In the present paper, this experiment is recalled only as a reference case, as it serves as a baseline for comparison with the two other reactor configurations investigated here. Accordingly, only the essential parameters are summarized below. The plasma reactor was an industrial coaxial plasma torch [39]; a photograph of a pure $CO_2$ plasma sustained at 220 W is shown in Figure 1. Plasma ignition occurred spontaneously at a frequency tuned to 2.49 GHz and at MW power levels between 180 and 200 W. During operation, the torch was surrounded by a quartz tube with an inner diameter of 26 mm, which is significantly larger than the transverse dimension of the plasma plume (see Figure 1).

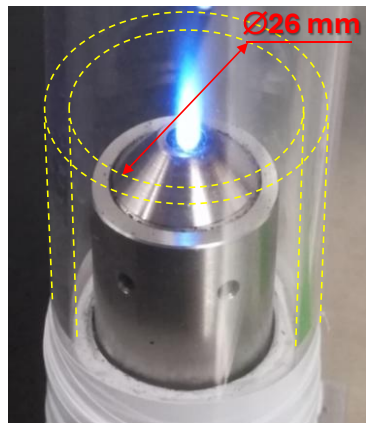

Figure 1. Photo of the $CO_2$ plasma discharge sustained with CW microwave power of ~200 W and 12 slm inflow in a coaxial plasma torch. Thin dashed lines sketch the surrounding quartz tube.

In that studies, the volumetric concentration of products in effluent, the time resolved rotational and vibrational temperatures as well as time resolved absorbed power in the plasma were measured [9, 26]. Additionally, time resolved fast imaging was utilized to follow a plasma evolution between two consecutive reignition events. Due to the converging geometry of the torch, the gas flow pattern is similar to a converging-diverging nozzle configuration [17], in which the gas flow first is accelerated in the nozzle neck and then decelerated in the diverging region. In the coaxial torch, the plasma is sustained at the top end of the torch and expands freely in the surrounding quartz tube, where it is quenched through convective mixing with cooler afterglow gas. The details of experiment and operation scenarios conducted in [9, 26] are summarized in Table 1.

Table 1. Summary of pulsed-operation scenarios in coaxial torch experiments at atmospheric pressure. More details can be found elsewhere in [9, 26].

| MW Power, W | $CO_2$ Flow, slm (at 25°C) | SEI, eV/molec | Typical Plasma Size, mm | Pulse Time, µs | Duty Cycle | Gas Supply | Afterglow & Quenching |
|---|---|---|---|---|---|---|---|
| 80÷250 | 12, 15, 18 | 0.02÷0.22 | h=10÷15 D=2.5÷3.5 | 0.1÷50.0 | 0.1÷0.99 | Axial flow between inner and outer conductor | Free expansion to surrounding tube ⌀26mm |

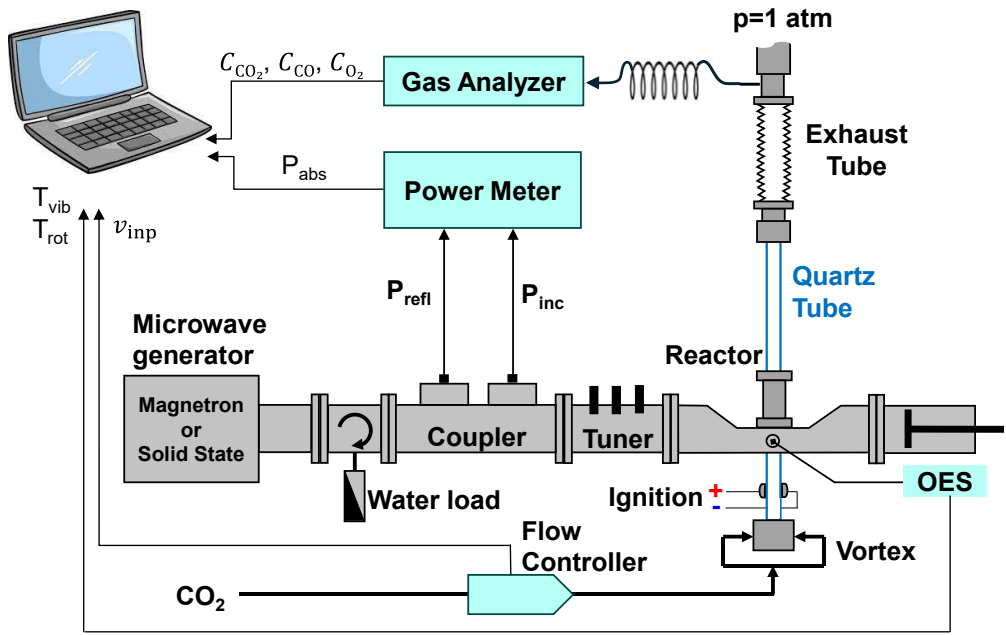

Figure 2. Schematic illustration of the Surfaguide-based reactor experiment.

## 2.2 Surfaguide-based reactor experiment (KIT)

The scheme of the Surfaguide-based reactor experiment is presented in Fig. 2. In this experiment, the Surfaguide-based [35] reactor was utilized to deliver MW power to a plasma confined in a quartz tube with an inner and outer diameters of 16 and 20 mm, respectively. As a part of the reactor, a chimney (a metallic tube surrounding the quartz tube with 110 mm in height) was installed to extend the coupling area of microwave to the plasma and to reduce the leakage of microwave to the surroundings. The plasma was ignited with the aid of a high-voltage pulse applied to the outer surface of the quartz tube at about 40 mm below the reactor and with argon flowing in the reactor tube. For high-voltage pulse generation, a standard automotive ignition-coil was utilized with the input voltage of 12 V. The output voltage was not explicitly measured but following the specifications for standard ignition-coils it is in the range 20 and 30 kV. After plasma ignition in argon, the argon flow was replaced by pure $CO_2$ (99.998%) gas flow. The exhaust tube was open to atmosphere, therefore the pressure in the reactor was equal to 1 atm or a bit higher because of some pressure drop in the heat exchanger installed in the exhaust line. The vortex configuration of the gas inflow was ensured by the geometry of the vortex injection system from Sairem. This system features two helically configured gas channels with an inner diameter of 2 mm. Such a vortex flow enabled efficient isolation of inner surface of reactor wall from the hot gas flowing near the axis of reactor tube. The amount of $CO_2$ gas supplied was controlled with a mass-flow controller. Next to the microwave generator output, the microwave circulator was installed to redirect and absorb the reflected wave and thus to minimize the secondary reflections inside waveguide system, that is an important condition for correct measurements of incident and reflected powers (Fig. 2). Between the circulator and reactor, a bidirectional microwave coupler (- 50 dB) served for decoupling the incident and reflected microwave for their subsequent detection.

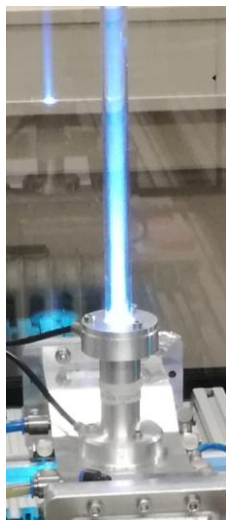

Figure 3. $CO_2$ plasma discharge sustained with pulsed microwave in the Surfaguide-based reactor. The $CO_2$ inflow rate is 15 slm and mean power is 2000 W.

For later discussion on the efficiency of $CO_2$ conversion in the above system, it is important to describe the plasma afterglow in more detail. Plasma discharge was confined in relatively long quartz tube whose length above the reactor amounted to about 600 mm. To cool the reactor tube, the dry air flow between chimney inner surface and tube outer surface was supplied, whose flow was approximately 100 slm and temperature between 21 °C and 23 °C. For the rest of quartz tube no forced cooling was applied, and surface-to-surrounding radiation and convective heat exchange between tube surface and surrounding atmosphere (21-23 °C) took place. The end of the quartz tube was sealed to a glass-to-metal transition unit which was water-cooled. Right after this transition a gas-decoupling element was installed to branch a portion of the off-gas toward gas analyzer, where $CO_2$, CO and $O_2$ relative volume concentrations were recorded as described in Section 2.4.3. The gas probe position was located about 620 mm from the center of the reactor. It is worth noting that the length of the plasma column depends on the mean microwave power supplied and the $CO_2$ inflow rate. The higher the mean power the longer the plasma column that agrees with results obtained at 900 mbar in [40]. At powers of > 2 kW the plasma column length can reach 30 to 40 cm (see photo in Fig. 3).

*2.3 Plasma torch at IPP with nozzle quenching*

Further experiments were conducted with a 2.45 GHz microwave cavity-based plasma torch (see Figure 4) operated in nozzle configuration. The nozzle configuration was chosen for these experiments as its quick quenching allows to achieve reasonably high energy efficiencies while keeping the effluent at temperatures high enough for waste heat reuse [41]. In another configuration with water-cooled effluent channels even higher $CO_2$ conversion (up to 57%) could be achieved [16], however at the cost of pushing the effluent temperature down to close to room temperature. The cavity of this microwave plasma torch, which was developed at University of Stuttgart [42, 43], consists of a combination of a cylindrical and a coaxial resonator. The plasma torch was operated in the same configuration that was used by R. Antunes et al. [41] in their oxygen separation experiments; that is the plasma was sustained in a flanged quartz tube with an inner diameter of 26 mm above which a water-cooled stainless-steel nozzle that was attached to the plasma reactor. The pressure in the plasma zone was set to 200 mbar for ignition, and was ramped up to 900 mbar immediately afterwards for the actual measurements. After passing through the nozzle, the product gas stream is cooled down to room temperature by flowing through a 2 m long heat exchanger. The composition of the product gas stream was analyzed at the end of the heat exchanger by means of mass spectrometry (see [44] for details on the mass spectrometry system). The gas was injected into the reactor via tangential gas inlets at the bottom of the coaxial resonator in order to create a vortex gas flow. Such a swirl gas flow protects the flanged quartz tube from melting by ensuring that the hot plasma gas is surrounded by layers of colder gas. The $CO_2$ flow rate in all experiments with this setup was set to 7.6 slm (at 25°C) and the pressure inside of the reactor after plasma ignition was kept constant at 900 mbar. At such a quasi-atmospheric pressure, the installation of a nozzle on top of the reactor was demonstrated to drastically enhance conversion and plasma energy efficiency of $CO_2$ conversion [45]. This performance enhancement is attributed to the quenching of the plasma off-gas imposed by the nozzle, because the hot gas in the center of the plasma reactor is forced to mix with the surrounding layers of colder gas and because of the contact of the hot gas to the water-cooled surfaces of the nozzle [2, 45, 46]. A microwave source used in actual experiment was the same solid-state microwave generator as one in Surfaguide-based experiment at KIT. A system of WR340 waveguides conducts the microwaves from the generator to the resonant cavity (see diagram at the top of Figure 4). A manual 3-stub tuner is utilized for impedance matching and power measurements are performed with the use of bi-directional cross-guide coupler and two peak power sensors.

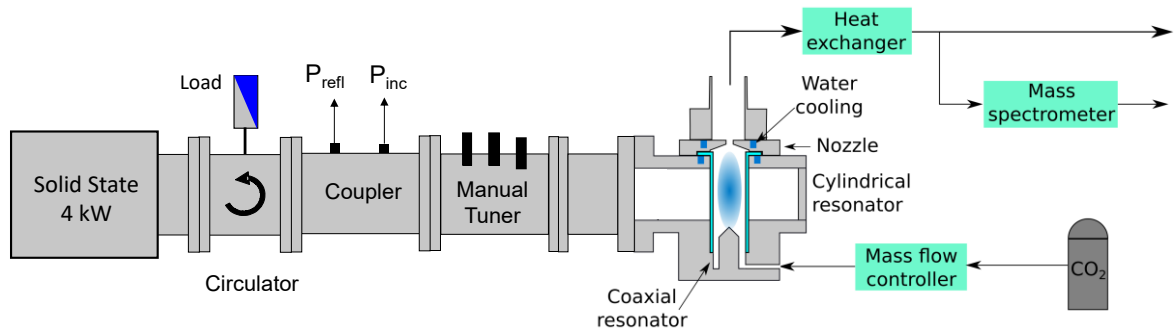

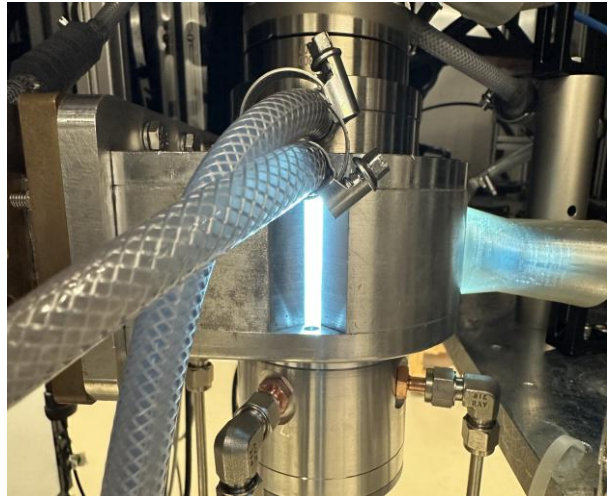

Figure 4. Plasma torch with the nozzle installed in IPP Garching: (top) schematics of the experiment with microwave and gas supplying and exhaust lines; (bottom) photo of operating torch in $CO_2$ splitting experiment.

*2.4 Methods*

*2.4.1 Time resolved temperature measurements*

In the coaxial torch and Surfaguide-based experiments, for temperature measurements of plasma species, an optical emission spectroscopy (OES) system synchronized with the microwave pulse generator was utilized. The synchronization in coaxial torch experiment was enabled by use of microwave detector measuring the pulse envelope of decoupled incident power [9] and in Surfaguide-based experiments by use of reference TTL pulse signal directly from SSM generator. Precise synchronization (< 1 ns) was achieved by choosing an appropriate delay in the digital delay generator (DDG) and allowed the use of accumulation mode for acquisition of plasma light emission. This significantly increases the signal-to-noise ratio and thus narrows the confidence interval for measured data. The use of the accumulation mode is an essential precondition at low photon's yield in a single acquisition with a short gate time. In iCCD camera Andor A-DH340-18U-03, repetitive acquisitions were summed up (*Integration-on-Chip*) at particular phase of pulse period that allowed to operate the iCCD camera with 50 ns gate. The OES system was wavelength-calibrated using a combined Hg/Ar/Ne spectral lamp and corrected for the relative spectral response of the optical detection system within the 420–580 nm spectral range. Plasma emission was collected using a collimating lens with an aperture of 4.0 mm. Accordingly, the transverse dimension of the light-collection region was approximately 4.0 mm. In the Surfaguide-based experiment, the collimating lens was mounted in an opening of the side wall of the reactor and aligned such that its line of sight was perpendicular to the reactor tube axis and directed toward the tube center. The collected plasma emission was transmitted through a 2 m optical fiber to the entrance slit of an Acton SP-2756 spectrograph. The rotational and vibrational temperatures were determined by fitting the C$_2$ Swan emission bands (for $T_{rot}$ >5000 K) and the CO Ångström emission bands (for $T_{rot}$<5000 K) using the MassiveOES code [47, 48]. The rotational temperature is assumed to be representative of the gas temperature due to rapid rotational-to-translational relaxation at atmospheric pressure [10]. It should be noted that the $T_{vib}$ derived from Ångström system corresponds to the vibrational population of the electronically excited CO(B$^1\Sigma^+$) state. Therefore, it represents an apparent vibrational temperature of the plasma and is used here primarily for comparative analysis between different operating regimes. The uncertainty for $T_{rot}$ and $T_{vib}$ is based on estimates in [10] and amounts to 500 K and 1100 K, respectively.

*2.4.2 Reflected and absorbed power measurements*

The accurate measurement of the absorbed power in the pulsed plasma enables not only the estimation of the process efficiency but also, if they made with sufficient time resolution, a qualitative interpretation of electron density evolution in the discharge [22, 25, 34, 49]. For the evaluation of both incident and reflected power in all three experiments, the real-time power sensors RTP5006 from Boonton Electronics were utilized [50]. These sensors allow to measure the power time profile with any pulse width >10 ns and rise/fall time >3 ns. Since the power pulse-form deviates from an ideal rectangular shape (especially in the reflected power signal) the power signals were recorded and integrated with MATLAB software to calculate the mean absorbed power as follows:

$$\overline{P_{abs}} = \tfrac{1}{T}\int_0^T P_{inc}(t)dt - \tfrac{1}{T}\int_0^T P_{ref}(t)dt \qquad (1)$$

Here, $\overline{P_{abs}}$ is a mean absorbed power, averaged over an entire period of power pulsation, and $P_{inc}$, $P_{ref}$ and $T$ are the incident power, reflected power and the period of power modulation $T=(t_{on} + t_{off})$, respectively. Note, that due to the group delay dispersion effect the form and effective width of incident pulse is changing while it propagates through the waveguide system and reactor. Therefore, the explicit integration in eq. 1 provides the correct estimation of mean microwave power in a pulsed regime. Additionally, the parasitic leakage of microwave power out the reactor, measured with aid of leakage detector, was always below 1 mW/cm² at about 5 cm away from reactor in all experiments. Therefore, it was neglected in the absorbed power estimation.

### 2.4.3 Measurement of product concentration in afterglow

In the coaxial torch and the Surfaguide-based experiment, effluent gas analysis was enabled with an X-STREAM XEGP gas analyser [51] from Emerson Electric. The gas travelling time from the reactor to the gas detecting sensor amounts ~5 seconds that enabled an ex-situ monitoring of gas composition versus time with a graphical user interface in the gas analyser. The gas analyser is equipped with detecting banks for measuring relative concentrations of $CO_2$, $CO$, and $O_2$ gases. The detection in the first two banks is based on non-dispersive infrared absorption and for oxygen concentration, the principle of paramagnetic resonance is utilized. In the coaxial torch and Surfaguide-based experiments, the concentrations of CO ($n_{CO,outlet}$), $CO_2$ ($n_{CO2,outlet}$) and $O_2$ ($n_{O2,outlet}$) in the off gas were monitored.

In the IPP plasma experiment, the gas composition in the afterglow was analysed with mass spectrometry (MS). The MS system consists of a spectrometer QMG 220 M2 Prisma Plus with cross-beam ion source from Pfeiffer Vacuum. It includes also a custom designed multi-stage gas sampling system enabling measurement of gas compositions over a wide pressure range [44].

### 2.4.4 Calculation of conversion and energy efficiency

To enable a comparable analysis of the $CO_2$ splitting performance in the different plasma reactors we calculate the specific energy input (*SEI*) based on the average absorbed microwave power in plasma ($\overline{P_{abs}}$) from equation 1. The *SEI* represents a mean amount of energy spent per $CO_2$ molecule and reads as follows:

$$SEI = \frac{\overline{P_{abs}} \cdot V_m}{F_{CO2}^{in} \cdot N_A}, \qquad (2)$$

where $F_{CO2}^{in}$, $V_m$, and $N_A$ are the gas inflow rate, molar gas volume (24.47 L/mol) and Avogadro constant, respectively. It should be noted that *SEI* is an average parameter because both absorbed power and axial gas velocity, $v_{gas\parallel}$ ($F_{CO2}^{in}$=cross-section area [m²] x $v_{gas\parallel}$ [m/s]) are considered as mean values over the cross-section of the reactor tube. Considering reaction stoichiometry in R1, the calculation of $CO_2$ conversion based on measured concentration of CO in off gas ($n_{CO,outlet}$) is performed as follows [52]:

$$\chi = \frac{n_{CO,outlet}}{1 - n_{CO,outlet}/2} \qquad (3)$$

In addition, the energy efficiency of the process is expressed as follows:

$$\eta = \chi \cdot \frac{\Delta H_R^0}{SEI} \qquad (4)$$

where $\Delta H_R^0$ is the enthalpy of the reaction R1.

## 3. Results and discussion

### 3.1 Temperature Measurements

#### 3.1.1 Temperature Measurements in Coaxial torch

For comparison reasons, we summarize in this section the main findings related to temperature measurements in the coaxial torch experiment [9, 26]. The microwave power pulsation leads to an effect that the plasma is essentially re-ignited with each new-coming energy pulse. Consequently, time resolved temperature measurements using optical emission spectroscopy (OES) were only possible during the microwave pulse and shortly thereafter, until the excited molecular states faded. The main finding is the discovery of nonequilibrium plasma state with $T_{vib}/T_{rot} \approx 2$ in the beginning of the pulse, provided the pulse time $t_{on} \geq 1.5$ µs (see Fig.10 in [26]). During the pulse, the rotational temperature rises nonlinearly from its initial value—typically between 3000 K and 4000 K—until it approaches the vibrational temperature, which spans between 7000 K and 8000 K. Once VT equilibrium is reached, both temperatures remain nearly constant throughout the remainder of the pulse and decrease rapidly after the pulse-end. It is worth noting that, that the increase of $T_{vib}$ in the beginning of pulse is not resolved within the 50 ns time resolution, therefore the e-V energy transfer is supposed to be faster than 50 ns. It is insightful that the time required for the thermalization is not constant but rather depends on pulsation timing. For scenarios with the same $t_{on}$=2.5 µs $T_{rot}$ recovers earlier in the scenario with shorter inter-pulse time $t_{off}$=3.75 µs as compared with scenario with longer $t_{off}$=7.5 µs (Fig.9 in [26]). The found approximate thermalization times of 800 ns (at $t_{off}$=3.75 µs) and 1600 ns (at $t_{off}$=7.5 µs) are at least in the same order of magnitude as the predicted VT relaxation time in [53]: for a given pressure and 3000<$T_g$<7000 K, it spans between 300 and 800 ns. As to the temperature decrease after pulse-end, the detection of the excited $C_2$ ro-vibrational bands was restricted because of their relative quick fading. Note, the period after pulse-end when the OES can still resolve reducing signal above the noise level indirectly reflects the rate of temperature decrease, and depends on pulsing scenario. It is hard to make a strong conclusion how the rate of energy dissipation from plasma is linked to the pulsation timing. Nonetheless, for two scenarios with the same $t_{on}$ (=2.5 µs), there is a hint that the cooling rate is lower for the scenario with a shorter $t_{off}$ (=3.75 µs) as compared with one with a longer $t_{off}$ (=7.5 µs); it is also indirectly confirmed by a shorter period detectability of excited bands after pulse-end: 650 ns and 1050 ns respectively (Fig. 9 in [26]).

| Scenario | $t_{on}$ | $t_{off}$ | DC | $\overline{P_{abs,pulsed}}$ |
|---|---|---|---|---|
| 1 | 0.2 µs, | 0.8 µs, | 0.20 | 0.75±0.02 kW |
| 2 | 5.0 µs, | 5.0 µs, | 0.50 | 1.88±0.10 kW |
| 3 | 5.0 µs, | 10.0 µs, | 0.33 | 1.26±0.10 kW |
| 4 | 10.0 µs, | 10.0 µs, | 0.50 | 1.84±0.10 kW |

Table 3. Pulse scenarios with different $t_{on}$ and $t_{off}$ times and fixed peak power of 4.2 kW, $CO_2$ inflow of 15 slm and atmospheric pressure.

#### 3.1.2 Temperature Measurements in the Surfaguide-based Reactor

In the Surfaguide-based reactor, energy coupling, plasma confinement, and gas flow configuration differ from those in the coaxial plasma torch. The plasma was sustained within a quartz dielectric tube which transverses the reactor, that results in a significantly larger plasma–microwave interaction volume as compared with the coaxial setup. This extended interaction region together with the efficient waveguide-based power delivery, enables plasma operation at kilowatt power levels. In the present pulsed power Surfaguide-based experiment, we kept the peak power at 4.2 kW and varied the duration of pulse ($t_{on}$) and inter-pulse ($t_{off}$) within a broad range of time intervals, as listed in Table 3. The reference CW scenarios were performed in such a way that mean absorbed power both in CW and pulsed regimes were approximately equal: $P_{abs,CW} \approx \overline{P_{abs,pulsed}}$. For rectangular pulses, the mean absorbed power can be also defined through the duty cycle (DC) as follows:

$$\overline{P_{abs,pulsed}} = DC \cdot P_{abs,peak} \qquad (5)$$

where, DC reads as

$$DC = t_{on}/(t_{on} + t_{off}). \qquad (6)$$

In the experimental setup, the pulse waveform is not perfectly rectangular, requiring $t_{on}$ and $t_{off}$ to be defined as the intervals between the two time points at which the power crosses 10% of its peak value. In the Surfaguide-based experiment showed empirically that for a given peak power of 4.2 kW the plasma can be sustained only in regimes with $t_{off}$ < 12 μs. For longer $t_{off}$ the plasma extinguishes.

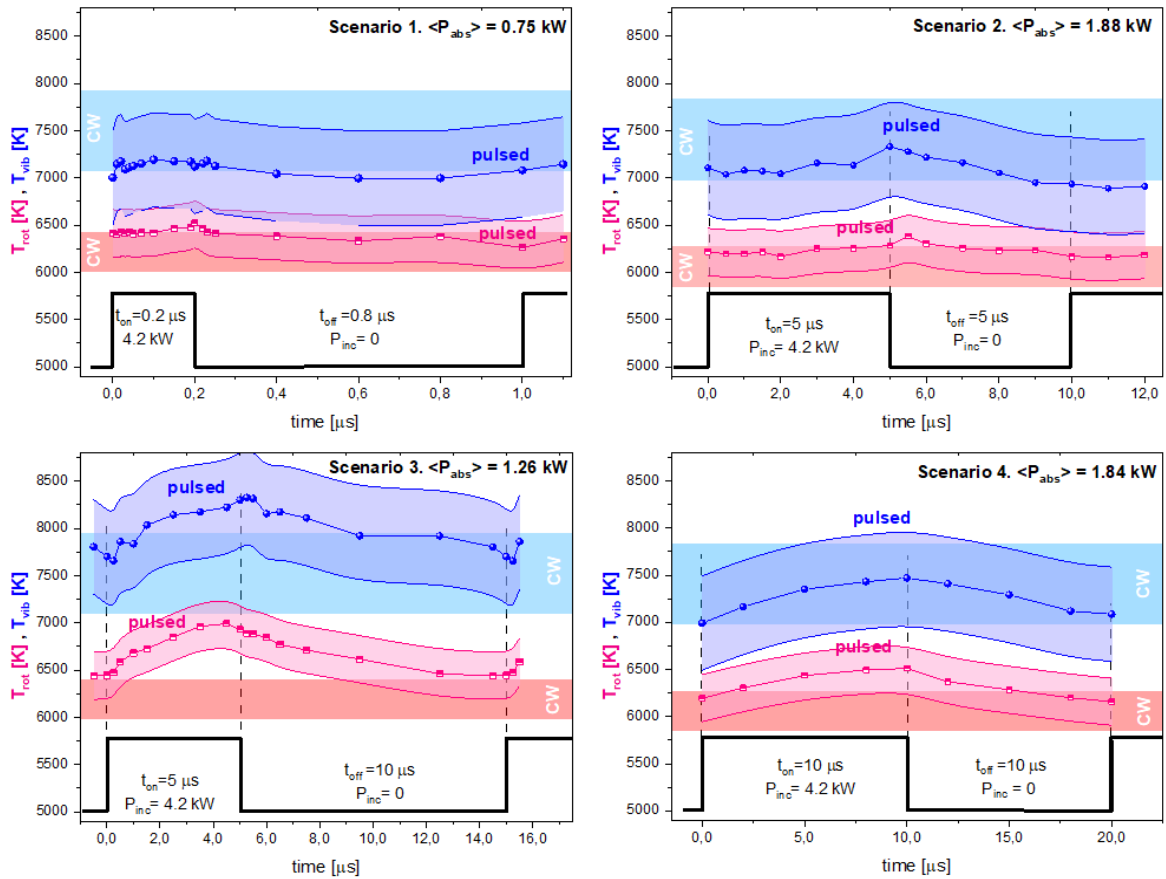

Figure 5. Rotational and vibrational temperatures measured in Surfaguide-based reactor for scenarios 1 to 4 listed in Table 3 with different $t_{on}$ and $t_{off}$ times. Results for CW scenarios are shown with horizontal light red and blue bars as a reference. Thick black lines sketch the timing for microwave power pulsations. Confidence interval is 500 K for $T_{rot}$ and 1100 K for $T_{vib}$.

Additionally, a lower limit of approximately 700 W was found for the mean absorbed microwave power required to sustain the plasma in the present setup. Note that the plasma ignition for all scenarios in the Surfaguide-based was always performed with CW microwave. To access the predefined pulsing scenario, the microwave generator is first switched to pulsed mode after plasma ignition with relatively large duty cycle and $t_{off}$ <12 μs, e.g. $t_{on}$ = 5.0 μs, $t_{off}$ = 5.0 μs. From that reference point, both $t_{on}$ and $t_{off}$ times were varied independently with the only restrictions that $t_{off}$ < 12 μs and $\overline{P_{abs,pulsed}}$ > 700 W. In this section we present time resolved temperature measurements performed in the four pulse scenarios as indicated in Table 3. The results are summarized in Figure 5. In scenarios 1 and 3 finer time steps were chosen with attempt to resolve fast events in temperature traces. Yet, the results have shown quite smooth time behavior both for rotational and vibrational temperatures. Consequently, in scenarios 2 and 4 the time steps were chosen coarser. In pulsed microwave plasma, the important difference between the Surfaguide-based and coaxial torch experiments is that the plasma in the Surfaguide-based does not extinguish between microwave pulses and therefore no reignition phenomenon occurs. Plasma, at least in the center of the Surfaguide-based reactor, exists both during power-ON and power-OFF time intervals. The evidence for that is the fact that the rotational $C_2$ Swan emission bands, which were used to derive plasma temperatures, are well detectable both in and out of microwave pulse. In Fig. 5 the rotational (red) and vibrational (blue) temperatures in pulsed 1-4 scenarios are presented. Additionally, temperatures in the continuous-wave (CW) regime are indicated by horizontal bars as a reference. A common trend observed for both rotational and vibrational temperature traces is an increase during the microwave pulse and a decrease during the inter-pulse period. This behavior qualitatively agrees with coaxial torch results [26], although the modulation of $T_{rot}$ in the coaxial torch was significantly larger—reaching up to 3500 K due to the reignition phenomenon. Both the increase and decrease are monotonic, with maximum amplitudes ranging from 100 to 500 K depending on the discharge conditions. Although the amplitude of the temperature modulation lies within the nominal resolution of the diagnostic, its presence is found to be statistically significant and phase-locked to the imposed power modulation [54, 55]. In scenario 3, with $t_{on}$ = 5.0 μs, $t_{off}$ = 10.0 μs, the temperature modulation is found at most pronounced. After new coming pulse, the gas temperature grows and reaches the level exceeding the minimum temperature recorded at end of OFF phase by about 500 K. Importantly, the amplitude of temperature increase during ON phase depends on $t_{off}$ – compare the scenario 2 and 3 with the same $t_{on}$ = 5.0 μs. The longer the inter-pulse time the bigger increase of temperature. The same is valid for scenarios 2 and 4 having the same DC (mean power) but different $t_{on}/t_{off}$ timings: 5.0 μs/5.0 μs (scenario 2) against 10.0 μs/10.0 μs (scenario 4). The longer OFF time in scenario 4 results in larger gas temperature modulation. Additionally, the mean gas temperature in pulsed operation exceeds that observed in the corresponding continuous-wave (CW) regime at the same average microwave power, thereby promoting more efficient gas heating. In this sense, power pulsation can be regarded as an effective control parameter for increasing the gas temperature beyond the level attainable under CW conditions. Assuming that thermal dissociation is the dominant mechanism governing $CO_2$ conversion at atmospheric pressure [12], the observed temperature increase can be directly correlated with the enhanced reactor performance (see section 3.2).

Why does the temperature during MW pulse surpasses CW temperature? Firstly, it is highly unlikely that the higher peak power in pulsation regime may result in an increase of plasma temperature. In accordance with results obtained by E. Carbone *et al* [10] where CW power was scanned between 0.9 kW and 2.7 kW and $CO_2$ plasma sustained at 0.925 bar pressure, the rotational temperature was about 6000 K independent on the supplied power. Our $T_{rot}$ data in CW scenarios with different microwave powers depicted in Fig. 5 also show nearly constant temperature of about 6300 K. On the other hand, the possible explanation for observed $T_{rot}$ and $T_{vib}$ variations might lie in the transient behavior of the electron component observed in pulsed plasmas with time resolved Thomson scattering diagnostic [20, 30]. After the end of microwave pulse, the electrons' population starts depleting and thus decreases to a level lower than the one during the pulse. In section 3.4 we will discuss that such variation in electron density can be proven through observed variation in reflected MW power. After end of pulse, electron temperature $T_e$ without driving electric field reduces and even faster than electron density [20, 30]. Consequently, the new-coming portion of microwave energy will be coupled to a lower number of free electrons as compared with CW mode, that leads correspondingly to gaining more microwave energy per one electron. This shifts electron energy distribution upward and leads to enhanced rates of inelastic processes, first of all, vibrational excitation and electronic excitation. The populating of vibrational states through e-V and V-V energy transfer is known to accompany by V-T relaxation though with slower rates than V-V, while electronic excited states are quenched through direct energy transferring into translational energy of heavier species. Those two mechanisms are the main channels enabling the gas heating in microwave plasma. The rate coefficients for inelastic processes depend nonlinearly (often strongly) on $T_e$. So, a higher $T_e$ in pulsed mode [18, 20, 31] deposits disproportionately more heat per unit energy than lower $T_e$ in CW mode, that can explain the observed higher rotational temperature during MW pulse. The relaxation of $T_{rot}$ after end of pulse is much slower than in coaxial torch experiment where reignition took place. After stopping the microwave energy at the end of the pulse, $T_{rot}$ relaxes to the level of about 6000 K as observed in CW plasma.

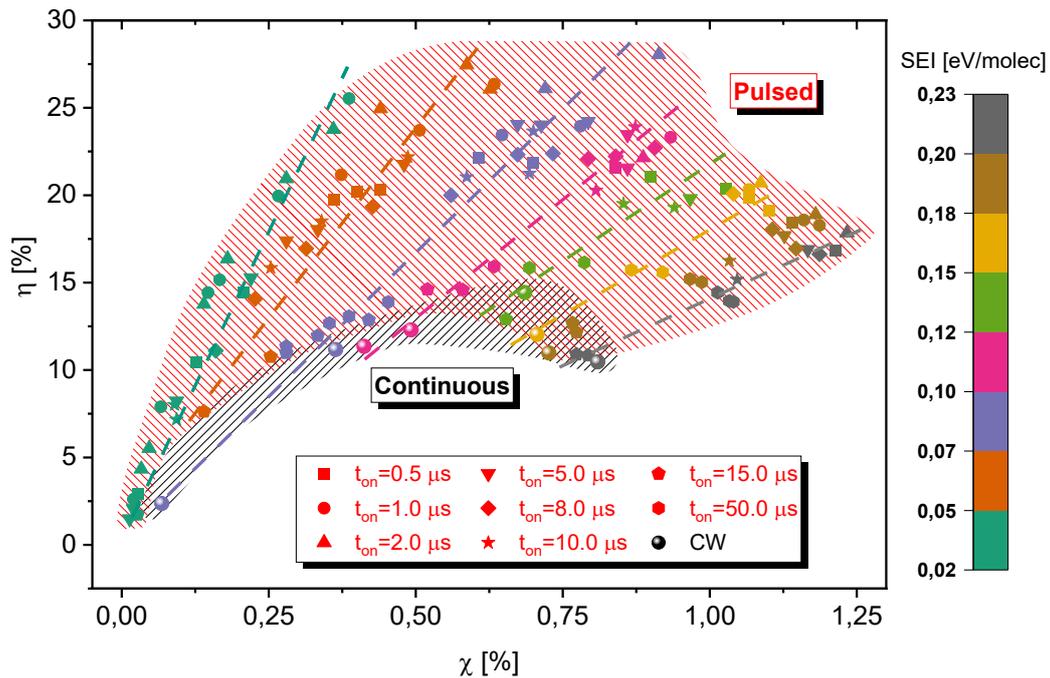

Figure 6. Energy efficiency ($\eta$) vs conversion ($\chi$) for pulsed and continuous wave (CW) plasma scenarios of $CO_2$ conversion in the coaxial torch. CW data are shown with spheres. Pulsed microwave data for different $t_{on}$ times (indicated on graph) are presented with different symbols accordingly. Hatched black and red areas indicate the ranges in ($\chi,\eta$)-space accessible in CW and pulsed scenarios respectively. SEI is colour coded. Dashed lines guide the points with SEI=const, which agrees with linear $\eta(\chi)$ dependency according to eq. 4.

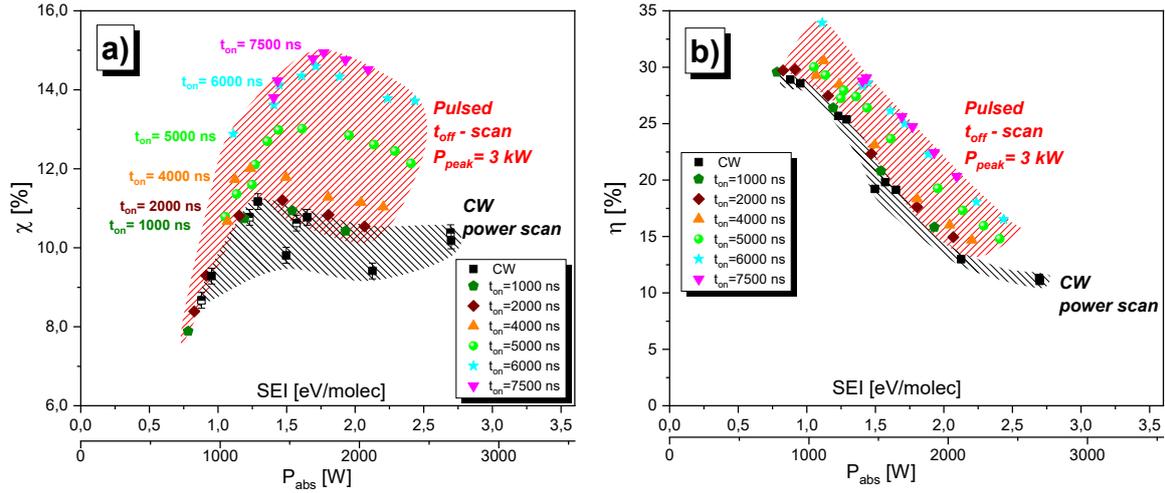

Figure 7. a) χ(*SEI*) and b) η(*SEI*) for plasmas in the Surfaguide-based reactor. Continuous wave (CW) data are shown with closed black squares as reference. Pulsed data for different $t_{on}$ times (indicated on graph) are presented with different symbols accordingly. Hatched black and red areas highlight assessed (χ, η) regions in CW and pulsed regimes respectively.

*3.2 Conversion and Energy Efficiency*

*3.2.1 Conversion and Efficiency in the Coaxial Torch*

As a reference, we present in this section a short summary of the results obtained in the coaxial plasma torch experiment published in [9, 26]. Estimation of conversion and energy efficiency for reaction R1 in the coaxial torch experiment demonstrated that the pulsed plasma regime outperforms the CW mode by a factor of two [9, 26]. The overall performance for all regimes with different pulsation times is presented as color-coded 2D plot in Fig. 6, where efficiency (η) is plotted over conversion (χ), and symbol's color represents specific energy input (*SEI*). This graph in fact summarizes the χ(*SEI*) and η(*SEI*) dependencies in Fig.1 a) and b) from [9]. The hatched black and red areas indicate the ranges in (χ,η)-space accessible in both CW and pulsed operation respectively. The dashed lines guide the linear dependence of η on χ at fixed *SEI* that agrees with eq. 4. More details on that experiment can be found in [9, 26]. The main message from that experiment is that the power pulsation advances the reactor performance both in χ and η by more than 100%.

*3.2.2 Conversion and Efficiency in the Surfaguide-based reactor*

In this conversion and efficiency estimation experiment in the Surfaguide-based reactor the peak power was kept constant at 3.0 kW. The pulse time ($t_{on}$) was varied from 1.0 μs to 7.5 μs. And for every fixed $t_{on}$, the $t_{off}$ time was varied gradually from smallest value up to the largest possible value while keeping minimum mean supplied power >700 W. For a given variations of $t_{on}$ and $t_{off}$ times, the *SEI* for all regimes spans within 0.7-2.7 eV/molecule. The obtained conversion and energy efficiency data were compared against reference CW plasmas sustained with different power levels (see Fig. 7a and 7b). The data are plotted versus *SEI* and absorbed power for convenience, with different symbols for every fixed $t_{on}$ regime. The CW data are represented by black squares and highlighted with black hatched area in Fig. 7 a and b. It is worth noting, that regimes with of $t_{on}$ = 1 μs and 2 μs reveal only minor improvements as compared with CW scenarios. At the same time, scenarios with $t_{on} \geq 4$ μs show remarkable progress in the reactor performance (highlighted with red hatched area). In all pulsation scenarios, the conversion and energy efficiency outperform the values obtained in CW plasma by a maximum of 40 % and 20 %, respectively at *SEI* values between 1.3 to 2.0 eV/molecule. The increase in the maximum reactor performance is reflected predominantly in the conversion rather than in the energy efficiency, in contrast to the coaxial torch, where the improvements in η and χ were more balanced. The maximum attained absolute conversion and energy efficiency amount to 15 % and 30 % respectively. In Figure 8, the χ(*SEI*) and η(*SEI*) dependencies are summarized in one (χ,η) color-coded plot. Scenarios with different $t_{on}$ times are shown with different symbols and *SEI* is color coded. The hatched black and red areas indicate the range in (χ,η)-space accessible in CW and pulsed scenarios respectively. The dashed straight lines for fixed *SEI* guide the eye to elucidate the linear dependence of η on χ that agrees with equation 4. The observed overall progress in pulsing regimes is smaller than one observed in coaxial torch, where relative increase amounted to ≥100% (see Fig. 6). Yet, the geometry of the two experiments and the *SEI* range are also quite different. In the coaxial torch the observed improvements referred to non-equilibrium states in the beginning of the pulse, reignition phenomenon and plasma volume increase, while in the Surfaguide-based, $T_{rot}$ and $T_{vib}$ are close to each other and are modulated not so significantly as in the coaxial torch. The observed performance enhancement of the Surfaguide-based reactor under pulsed operation can be attributed to the influence of power modulation on the electron dynamics, which is discussed in more detail in section 4.

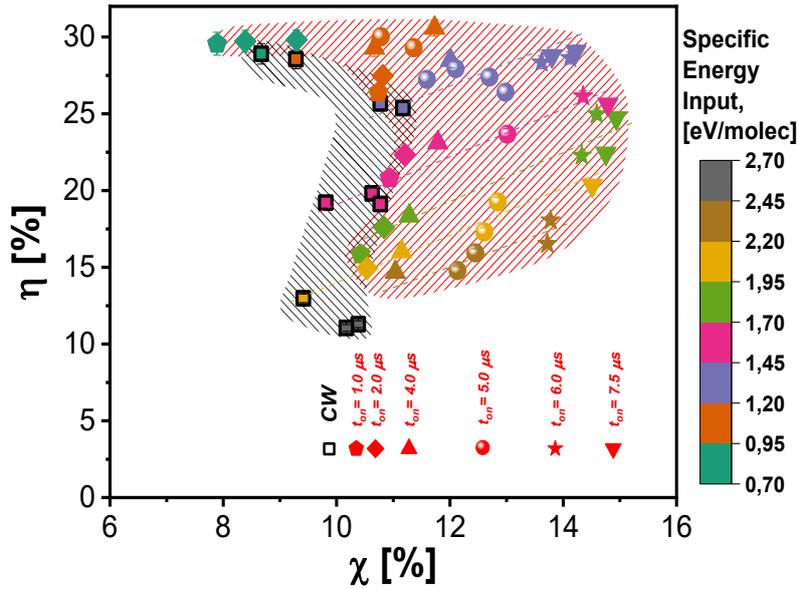

Figure 8. Energy efficiency versus conversion for $CO_2$ splitting in Surfaguide-based reactor for investigated regimes. Continuous wave (CW) data are shown with squares. Pulsed data for different $t_{on}$ times (indicated on graph) are presented with different symbols accordingly. Colour coding corresponds to specific energy input (*SEI*). Hatched black and red areas highlight assessed ($\chi, \eta$) regions in CW and pulsed regimes respectively.

## 3.2.3 Conversion and Efficiency in the IPP Plasma Experiment

In this section, we report on the $CO_2$ conversion rates and energy efficiency in the third investigated plasma reactor where pulsed microwave power was applied. Among all three considered configurations, only the plasma torch installed at IPP is equipped with an active gas temperature quenching in the reactor afterglow. As in the two other reactors, we have used also here the continuous wave (CW) scenario as a reference to elucidate the changes in conversion and efficiency of the process when CW is replaced with pulsed energy supply. For variation of specific energy input we have varied 1) the microwave input power in the CW scenario, 2) the inter-pulse time, $t_{off}$, at fixed $t_{on}$ time in the pulsed scenario. The following $t_{on}$ pulse times were considered: 1.0 μs and 4 μs. Note, because of lower $CO_2$ input flow (7.6 slm) than one in the Surfaguide-based experiment, the $SEI$ spans from 1.5 to 5.0 eV/molecule, while in the Surfaguide-based experiment it was within 0.7 – 2.7 eV/molecule.

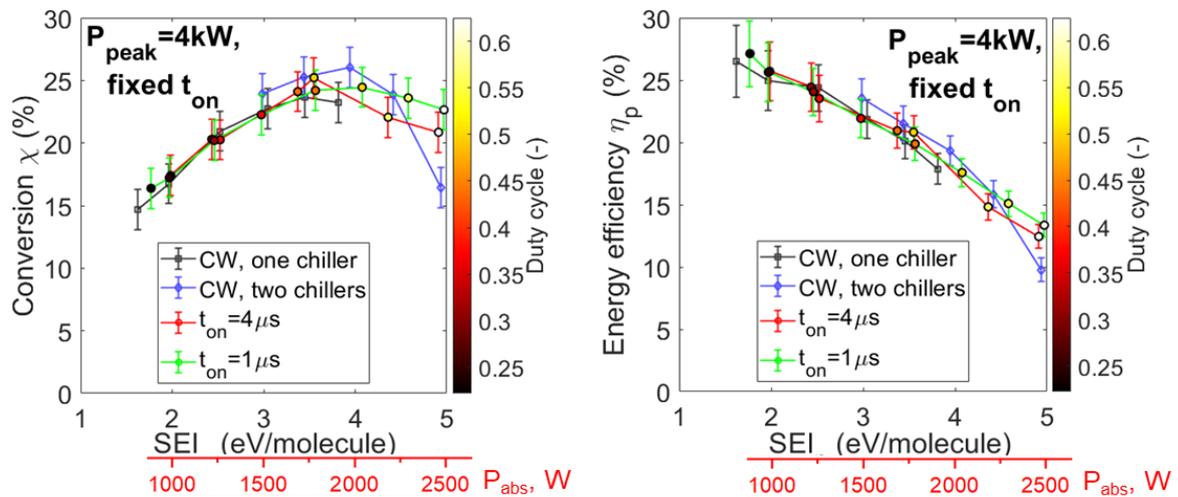

Figure 9. Plasma torch at IPP: (a) $CO_2$ conversion vs $SEI$ and absorbed microwave power. Continuous wave (CW) data shown with blue and black lines. Pulsed microwave data for $t_{on}$=1 μs and 4 μs are presented with green and red. Parameters: p= 0.9 bar, $F_{CO2}$=7.6 slm, $P_{peak}$=4 KW, $D_{in}$=26 mm

The results are summarized in Figure 9 where the conversion and efficiency are plotted against $SEI$ and absorbed microwave power ($P_{abs}$). Additionally, the symbols are filled with colors, that are coded according to the duty cycle ($DC$) values which are proportional to the mean supplied power. Note that at higher CW powers (correspond to higher $SEI$ on that graphs), the cooling power of one chiller was not enough and two chillers in parallel were utilized – see blue and black circles in Fig. 9. Qualitatively, the conversion has the similar peaked behavior as in the Surfaguide-based experiment (see Figure 7 a) – yet, the peak is at around 3.5 eV/molecule while the maximum in the Surfaguide-based data, depending on pulsation regime, lies between 1.24 and 1.80 eV/molecule. Notably, no difference between CW and pulsed data as well as between two pulsation regimes is detected. Similar to the Surfaguide-based experiment, energy efficiency demonstrates the same falling trend along with the $SEI$. Also here, no difference between pulsed and CW regimes is found (Figure 9 b). In the next experiment, to investigate the pulsed regime in a broader parameter range, we have varied both $t_{on}$ and $t_{off}$ times but keeping duty cycle constant ($DC$≈0.437) that corresponds to $SEI$=3.5 eV/molecule where maximum in χ($SEI$) data is detected. The resulting conversion and energy efficiency are plotted in Figure 10 versus $t_{on}$ time [56]. Note, as the $DC$ was fixed, the progression along $t_{on}$ axis corresponds to an increase in the corresponding $t_{off}$ values. The obtained data, also here, show no difference between CW and pulsed data within the confidence interval. Also, no dependency on $t_{on}$ (correspondingly $t_{off}$) time was found both for conversion and efficiency. It is worth noting that at all experiments with a given peak power of about 4 kW, the maximum $t_{off}$ time when plasma was relatively stable was up to 10.3 μs. At longer $t_{off}$ times, plasma became unstable and these results are not analyzed in present paper. At the same time, the found limit is close to maximum $t_{off}$ = 12 μs, observed in the Surfaguide-based experiment.

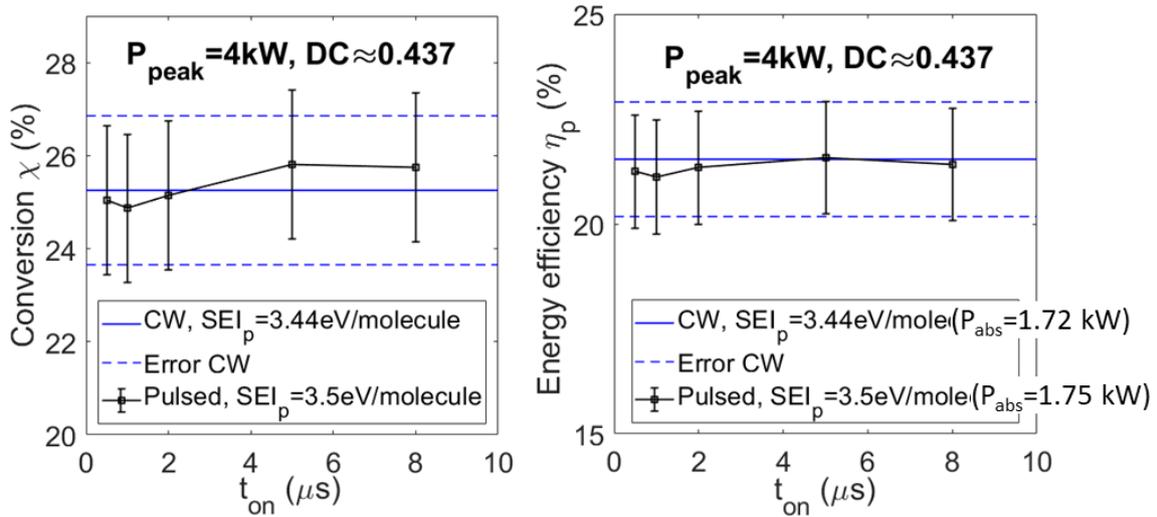

Figure 10. Plasma torch installed at IPP. Energy efficiency and conversion for different pulse times ($t_{on}$) at fixed $DC$=0.437 [56].

The explanation for observation, that no difference was found between CW and pulsed operation, which is in contradiction to the results obtained in the Surfaguide-based experiment, can lie in the difference of afterglow treatment. The gas cooling trajectory in the Surfaguide-based reactor and in the plasma torch at IPP are significantly different. The advanced quenching system in the afterglow of the plasma torch at IPP enables intense cross mixing and quenching of the hot gas in the water-cooled nozzle that weakens back reaction pathways significantly. By comparing the conversion in CW operation for $SEI \approx 2$ eV/molecule in Fig. 7 a and 9 a we may see that the rapid quenching in the plasma torch at IPP nearly doubles the rate from ≈9% to ≈17%.

In the Surfaguide-based experiment, the gas quenching is rather weak and slow. Effectively, the hot region where the dissociation reactions take place is extended into after-reactor zone. It can be evident also from visual observation of plasma in Surfaguide-based reactor in Fig. 3 where bright, intense light emitting plasma area expands a lot above the reactor. In modelling work of V. Vermeiren and A. Bogaerts it is shown that especially for high $SEI$ when the residence reactor time is relatively short, the temperature and the conversion may continue to rise in the afterglow region [57]. We suppose that plasma in the Surfaguide-based reactor takes maximum advantage from the increase of gas temperature in the pulsed regime due to its extended effective residence time (slow and late quenching) as compared to IPP experiment. In the same work V. Vermeiren and A. Bogaerts have shown, that the earlier quenching, before maximum conversion is reached, can be counter-productive [57]. In summary, the system with very fast and massive quenching in the water-cooled nozzle as in the plasma torch at IPP, apparently, cannot benefit from pulsation scenario because of its shorter effective residence time, as compared with the Surfaguide-based reactor under consideration.

*3.3 Reflected power as a monitor for electron density*

In this chapter, we consider how the reflected power signal can be related to the dynamics of electron density in power pulsation scenarios. The time-averaged electromagnetic power absorption density (Joule dissipation) in a conductive medium reads as follows:

$$p_{abs} = \frac{1}{2} \cdot Re[\sigma]|E|^2 \qquad (7)$$

where $\sigma$ and $E$ are the conductivity of the plasma and the electric field of microwave. For harmonically oscillating electric field with angular frequency $\omega$, the conductivity of cold collision plasma is expressed as follows [58]:

$$\sigma = \frac{e^2 \cdot n_e}{m_e \cdot (v_{en} - i \cdot \omega)} \qquad (8)$$

where $n_e$, $e$, and $v_{en}$ are electron density, electron charge and electron-neutral collision frequency. From eq. 7 and 8, it results in:

$$p_{abs} = \frac{1}{2} \cdot \frac{e^2 \cdot n_e \cdot v_{en}}{m_e \cdot (v_{en}^2 + \omega^2)} |E|^2. \qquad (9)$$

Assuming Maxwellian electron energy distribution function with electron temperature $T_e$, the electron-neutral particle collision frequency can be expressed as follows [59]:

$$v_{en}(t) = n_n \cdot \overline{\sigma_{mt}}(T_e) \cdot \sqrt{\frac{8 \cdot k_B \cdot T_e}{\pi \cdot m_e}}, \qquad (10)$$

where $n_n$, $\overline{\sigma_{mt}}(T_e)$ and $k_B$ are neutral particle density, mean electron momentum transfer cross section (which depends on $T_e$) and Boltzmann constant respectively. To make a transition to a full absorbed power, one needs to integrate equation 9 over the volume where the microwave field is coupled into the plasma. Considering a maximum electron density of approximately $1 \times 10^{19}$ m$^{-3}$ for given plasma conditions [60] together with the electron-neutral collision rate at atmospheric pressure, the estimated microwave penetration depth at 2.45 GHz is on the order of 10 mm. This indicates, that plasma can be treated like effectively transparent lossy medium and above integration over the volume is well justified. Accordingly, temporal variations in the spatial distributions of the electron density $n_e$ and electron temperature $T_e$ are assumed to be less significant than the evolution of their volume-averaged values, defined as: $\langle n_e \rangle(t) = \frac{1}{V} \int_V n_e(x,y,z,t)$ and $\langle T_e \rangle(t) = \frac{1}{V} \int_V T_e(x,y,z,t)$. In addition, we define the volume-integrated electric field energy as follows: $W_E(t) = \frac{1}{2} \int_V |E(x,y,z,t)|^2$. Under these approximations, and considering that the electron-neutral collision frequency significantly exceeds the microwave angular frequency ($v_{en} \gg \omega$) at 2.45 GHz, the eq. 9 can be reformulated to express the instantaneous absorbed power as:

$$P_{abs}(t) = \frac{e^2}{m_e} \cdot \frac{\langle n_e \rangle(t)}{\langle v_{en} \rangle(t)} \cdot W_E(t), \qquad (11)$$

where the effective electron-neutral collision frequency is given by

$\langle v_{en}\rangle(t) = n_n(t) \cdot \overline{\sigma_{mt}}(\langle T_e\rangle(t)) \cdot \sqrt{\frac{8 \cdot k_B \cdot \langle T_e\rangle(t)}{\pi \cdot m_e}}$. Here, the neutral density $n_n(t)$, because of weak pressure variations, is primarily governed by the gas temperature ($n_n \sim 1/T_{gas}$) whereas the mean cross-section for momentum transfer $\overline{\sigma_{mt}}$ can vary significantly as the mean electron temperature $\langle T_e\rangle(t)$ is expected to be modulated significantly along with modulated MW power. As a discharge volume and microwave coupling efficiency are time dependent, the variation of electric field energy integral $W_E(t)$ can also contribute to the overall absorbed power. Taking these effects into account, the instantaneous reflected power can be expressed as:

$$P_{ref}(t) = P_{inc}(t) - \frac{e^2}{m_e} \cdot \frac{\langle n_e\rangle(t)}{\langle v_{en}\rangle(t)} \cdot W_E(t). \tag{12}$$

Since the incident power $P_{inc}(t)$ is known, eq. 12 provides a direct link between the temporal evolution of the measured reflected power and the dynamics of the volume-averaged electron density $\langle n_e\rangle(t)$, the effective electron-neutral collision frequency $\langle v_{en}\rangle(t)$, and the electric field integral $W_E(t)$ in pulsed plasma operation.

In the following two sections, the behavior of the reflected microwave power signal is compared for two distinct regimes: one exhibiting plasma reignition (as observed in the coaxial torch reference case) and one without reignition (as in the Surfaguide-based and cavity-based plasma torches). It will be shown that, in addition to plasma temperature, the instantaneous reflected power serves as a sensitive indicator of plasma evolution during the power-OFF and power-ON phases.

### 3.3.1 Reflected power in coaxial torch experiment

As a baseline for comparison with the two kilowatt-range microwave experiments, this section examines the temporal behavior of the reflected power in the coaxial plasma torch operated under pulsed conditions. In this reference case, the $CO_2$ flow rate was 12 slm, while the peak microwave power was limited to 250 W. It should be noted that the impedance of the transmission lines – both between the microwave source and the plasma, and between the source and the power detectors – was kept constant throughout the experiments. Therefore, any observed impedance mismatch can be attributed solely to changes in plasma properties. Figure 11 presents the reflected power signals for three discharge conditions with identical pulse duration ($t_{on}$ =2.0 μs) but different inter-pulse times ($t_{off}$=12.0, 6.0 and 1.5 μs). All traces exhibit a pronounced peak at the beginning of each pulse. This initial peak in reflected power (Fig. 11 b) reflects a transient competition between the rapidly rising incident power (due to the initial impedance mismatch) and the nascent plasma whose volume [9] and $\langle n_e \rangle$ increase over time. In other words, the simultaneous increase in $\langle n_e \rangle$ and in the factor $W_E$ (associated with the expanding discharge volume and improving microwave coupling) leads to a progressive rise in the plasma absorption capability. According to eq. 12 this absorption scales as $\langle n_e \rangle(t)/\langle \nu_{en} \rangle(t) \cdot W_E(t)$ and once it becomes comparable to or exceeds $P_{inc}(t)$, a corresponding decrease in the reflected power is observed. Thus, in the reignition regime, the characteristic peak structure in the reflected power arises from the temporal interplay between the rising pulse profile of $P_{inc}(t)$ and the increasing absorption capacity of the developing plasma. The difference in timing and amplitude of the observed maxima for different inter-pulse times (see Fig. 11 b) can be attributed to variations in seed ionization, which seems to be the lowest for the longest inter-pulse time $t_{off}$. Correspondingly, the lower rate of seed ionization leads to a later breakdown and slower plasma development, that makes the first term in eq. 13 more dominating and leading to a peak with higher amplitude and later time coordinate: 21 ns, 23 ns and 30 ns for $t_{off}$ = 1.5 us, 6.0 μs and 12.0 μs respectively. As to dependence of power absorption on electron-neutral collision frequency, it scales with electron and gas temperatures as $1/T_{gas} \cdot \overline{\sigma_{mt}}(T_e) \cdot \sqrt{T_e}$ (see eq. 11). The time scale for $T_{gas}$ is much slower than rising time of microwave therefore $T_{gas}$ in the beginning of pulse can be considered nearly constant: ≈3000 K [9]. Yet, the variation of $T_e$ whose dynamic in pulsed plasma is a way faster than $n_e$ [61] cannot be neglected. Supposing the possible increase of $T_e$ from 0.2 eV to 1.0 eV (by extrapolating results in [61] for a given discharge conditions) and assuming Maxwell electron energy distribution, $\nu_{en}(t)$ increases approximately from 2E11 s$^{-1}$ to 4E11 s$^{-1}$. It results in a falling trend of $1/\nu_{en}(t)$ which may contribute only to the reduction of $P_{abs}$ and increase of $P_{ref}$ (see eq. 13). Therefore, the expected elevation of $\nu_{en}(t)$ due to increasing $T_e$ in the beginning of pulse does not change qualitatively the discussed above dynamic of $P_{ref}(t)$: by hampering the absorption, the $\nu_{en}$ can only contribute to the amplitude of $P_{ref}$ peak and its shift to a later time coordinate. Thus, among two fast responding factors, $n_e$ and $\nu_{en}$, the considered peaked behavior of reflected power in the beginning of pulse is supposed to be mostly driven by mean electron density evolution. At later times in the microwave pulse, $P_{inc}$ is constant and a discharge enters a quasi-steady state, when $\langle T_e \rangle$ and $\langle n_e \rangle$ are also supposed to be constant. Some increase in the reflected power can be attributed to the possible change in $n_e$ spatial distribution that may decrease the effective coupling volume and/or microwave matching quality – both lead to decrease of factor $W_E$ in eq. 13. Note, the time coordinates of that minimum: 33 ns, 74 ns and 1.39 us depend on $t_{off}$ and correspond to $t_{off}$ = 1.5 us, 6.0 us and 12.0 us, respectively (Fig. 11 b). The observed different timing of minimum position can be explained, by the transient competition of still growing mean electron density $\langle n_e \rangle$ and microwave mismatch due to the change of $n_e$ profile. Effectively, the position of this minimum in $P_{ref}(t)$ can be correlated with an ending time for $n_e$ building-up phase. It is worth noting that at the end of pulse the reflected power jumps one more time (see Fig. 11 a), yet this increase is supposed to be related to the pulse propagation physics, which will be discussed in more detail in the next section.

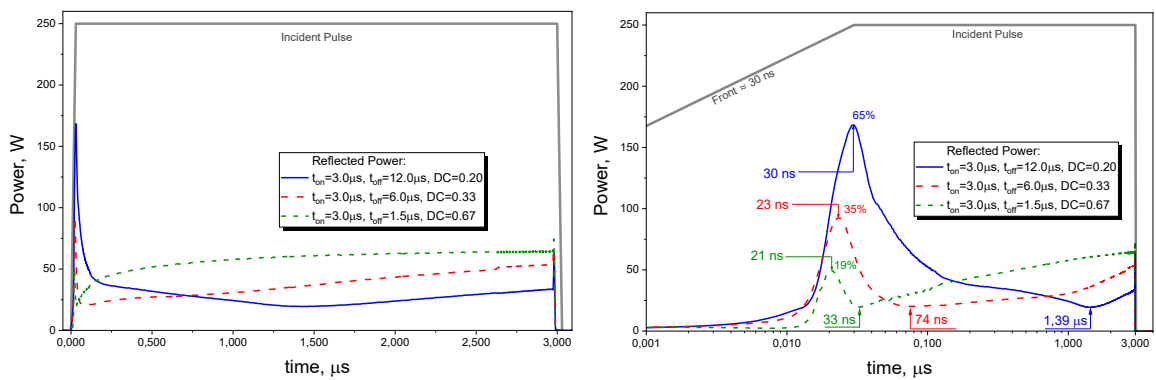

Figure 11. Reflected power in coaxial torch experiment (plasma in reignition regime) with $t_{on}$ =2.0 μs and $t_{off}$ = 12.0, 6.0 and 1.5 μs, or $DC$ = 0.20, 0.33 and 0.67 respectively. Incident pulse is shown schematically with thick grey solid line. a) Linear scale. (b) Logarithmic scale to elucidate the evolution at the beginning of pulse. The times and amplitudes (percentage of incident power) for the first maximum are indicated as well as timing for minimum position.

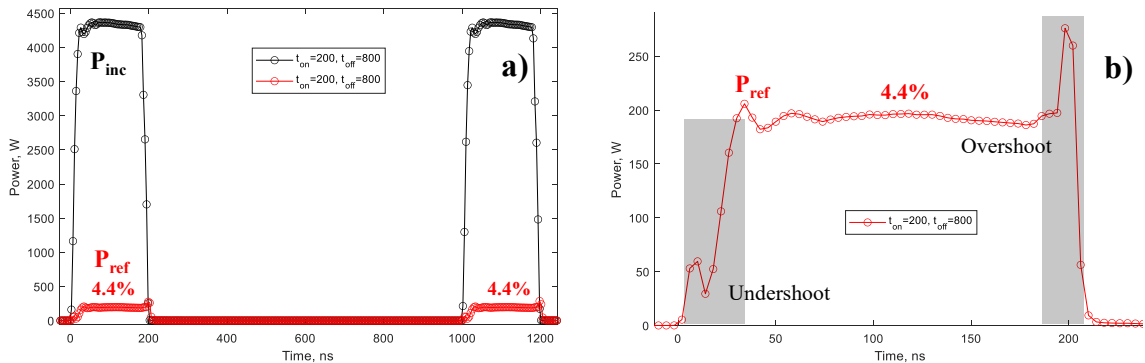

Figure 12. Time traces for incident and reflected microwave power measured in Surfaguide-based reactor (no reignition) with $t_{on}$ =200 ns and $t_{off}$ =800 ns (DC=0.20) and peak power of 4.2 kW. (a) both $P_{inc}(t)$ and $P_{ref}(t)$ on a time span of 1300 ns, (b) $P_{ref}(t)$ on a time span of 220 ns. Mean level of $P_{ref}$ in percentage of incident power is labeled.

### 3.3.2 Time resolved reflected power in Surfaguide-based reactor and plasma torch at IPP experiments

In two kilowatt-range pulsed plasma experiments, Surfaguide-based reactor and the cavity-based plasma torch, the incident and reflected powers were measured using the same pulse power detectors as in coaxial-torch experiment. For energy decoupling from the main transmission line, a 50 dB bi-directional cross-guide coupler was utilized. The rising and falling edges of the incident power exhibit characteristic times of approximately 15 ns. Figure 12 and 13 present the time traces for the incident and reflected power for operating conditions with $t_{on}$=200 μs, $t_{off}$=800 μs, $P_{peak}$=4.3 kW and with $t_{on}$=500 μs, $t_{off}$=650 μs, $P_{peak}$=4.07 kW, for the Surfaguide-based reactor and cavity-based plasma torch, respectively At the onset of the pulse, the reflected power exhibits a small perturbation, appearing as an overshoot in the cavity-based torch (Fig. 13b) and as an undershoot in the Surfaguide-based experiment (Fig. 12b), followed by steady-state levels of approximately 0.5% and 4.4% of the incident power, respectively. The overshoot amplitude in the cavity-based torch is relatively low, reaching about 1.2% of the incident power. In contrast, the initial peak observed in the coaxial plasma torch is significantly larger, ranging from approximately 20% to 65% of the incident power depending on the inter-pulse time $t_{off}$ (see Fig. 11). In addition, a ringing feature at the beginning of the pulse is observed in both the incident and reflected power signals in the cavity-based torch. This behavior is likely associated with internal reflections within the waveguide structures, leading to interference effects. At the end of the microwave pulse, an overshoot is observed in the reflected power signals for both experimental configurations. The key difference compared to the reference coaxial torch is the absence of a pronounced spike at the beginning of the power-ON phase. In the coaxial case, the amplitude of this spike strongly depends on $t_{off}$, indicating a plasma re-ignition regime. In contrast, such behavior is not observed in the kilowatt-scale systems studied here. This suggests that a relatively high residual electron density persists during the power-OFF phase, enabling a smooth transition between power-OFF and power-ON states. This interpretation is further supported by the observation that the plasma remains well-developed throughout the entire modulation cycle (see Section 3.1.2). It is worth noting that a similar overshoot feature in the reflected power signal is also observed at the end of the pulse, exhibiting comparable amplitude and temporal width. Importantly, the amplitude of these spikes appears to be independent of the power-OFF time. This suggests that the observed perturbations are likely associated with wave propagation effects rather than intrinsic plasma dynamics. Components such as the 3-stub tuner, couplers, bends, and waveguide-to-coaxial adapters can induce the reflections and standing wave patterns if the impedances are not perfectly matched and the system attenuation is low. To distinguish this from plasma effects on the reflected power, the microwave pulse propagation was simulated in a waveguide system with geometry relevant to real experiments. In Figures 14 and 15, the simulated geometries for both plasma experiments consisting of H- and E-bends, 3-stub tuner, 50 dB bi-directional cross-guide coupler and plasma reactors are presented, respectively. The full-wave 3D simulations were performed with the aid of CST Studio Suite package from Dassault Systèmes [62]. The plasma column was simulated as the nested cylindrical objects each of which has the constant plasma permittivity ($\varepsilon'_p = 1 - \omega_p^2/(\omega^2 + \nu_{en}^2)$) and loss factor ($\varepsilon''_p = \omega_p^2 \cdot \nu_{en}/(\omega^2 + \nu_{en}^2)$), where $\omega_p = 56.4\sqrt{n_e}$. The electron density profiles $n_e(r)$ used in the model were manually constructed as stepwise functions for each experimental configuration, following a peaked radial distribution. The central electron density, $1 \times 10^{19}$ m$^{-3}$, was based on interferometric measurements reported for similar conditions in [62], while the remainder of the profile was prescribed, with peripheral densities as low as $5 \times 10^{15}$ m$^{-3}$.

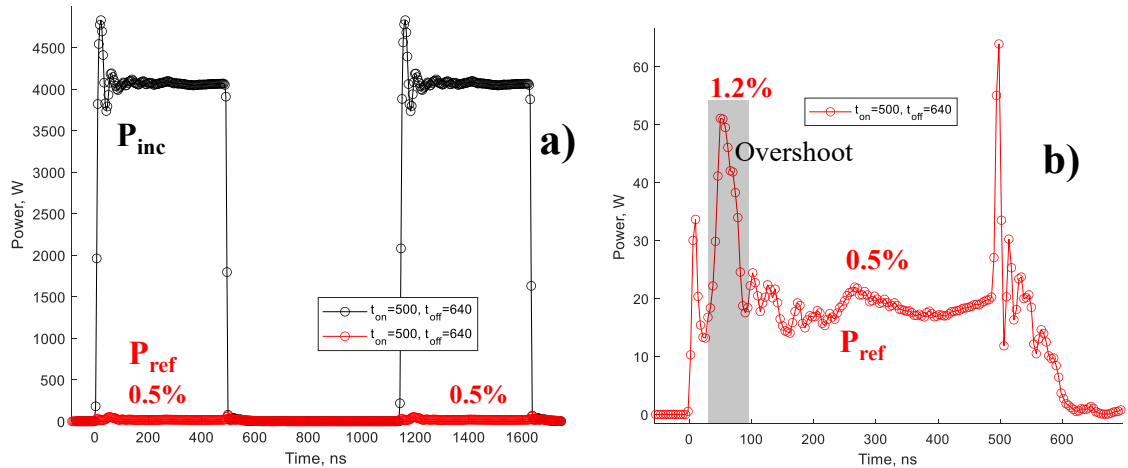

Figure 13. Cavity-based plasma torch at IPP experiment with $t_{on}$ =500 ns and $t_{off}$ =640 ns (DC=0.437) and peak power of 4.07 kW and 7.5 slm CO2 inflow. (a) Time evolution of incident (black) and reflected (red) microwave power; (b) time trace of reflected power at time span of 700 ns. Mean level of $P_{ref}$ in percentage of incident power is labeled.

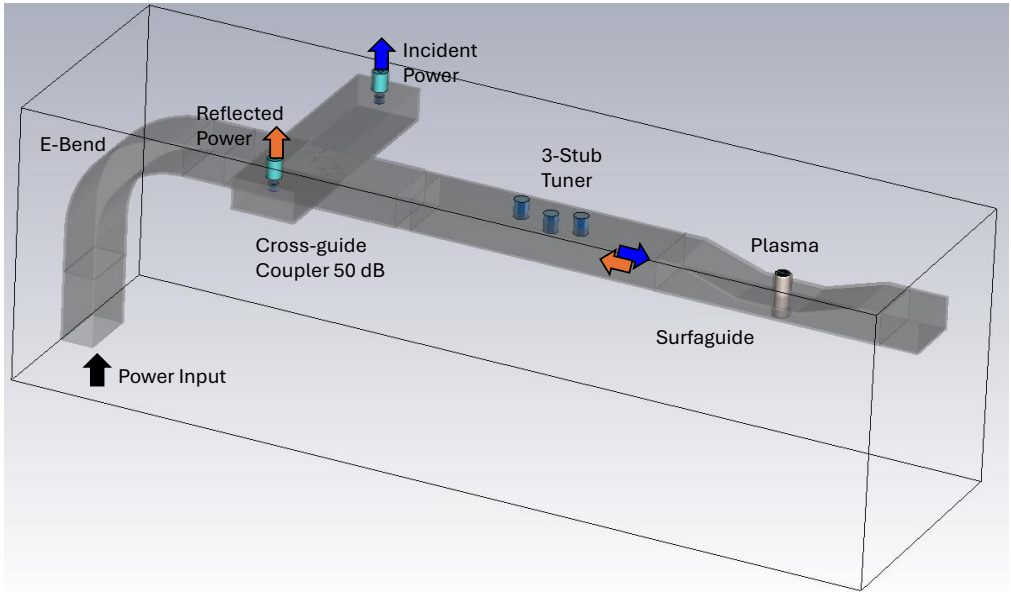

Figure 14. Full-wave 3D model for microwave propagation within waveguide system of Surfaguide-based experiment including bends, 3-stub tuner, 50 dB bi-directional coupler and plasma reactor. Black and blue arrows indicate the forward propagation whereas orange arrow - back propagating wave.

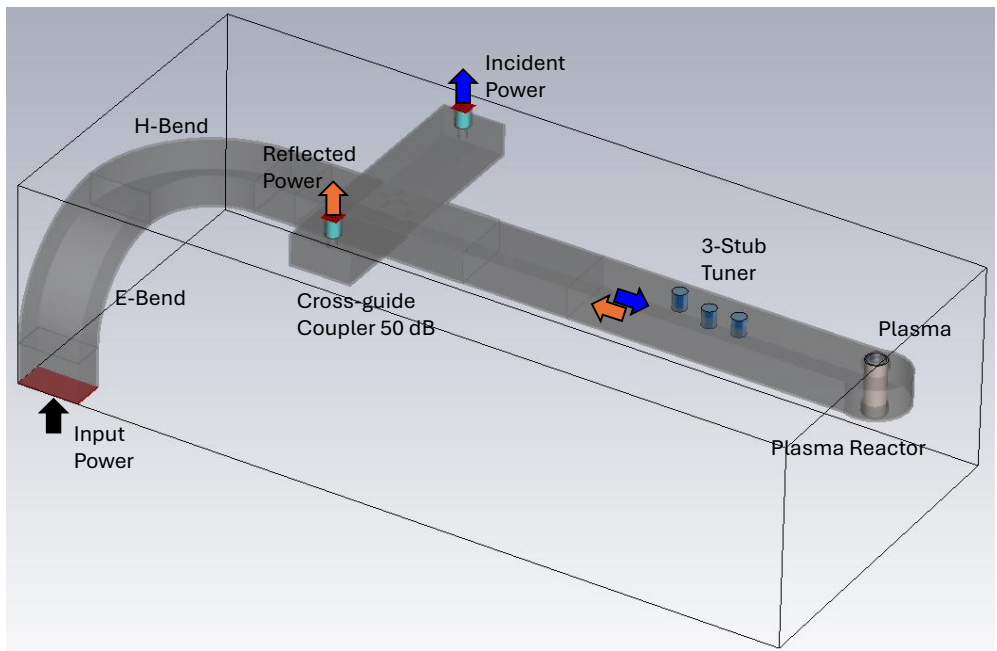

Figure 15. Full-wave 3D model for microwave propagation within waveguide system in IPP Garching experiment including E- and H-bend, 3-stub tuner, 50 dB bi-directional coupler and plasma reactor with quartz tube and plasma inside. Black and blue arrows indicate the forward propagation whereas orange arrow back propagating wave.

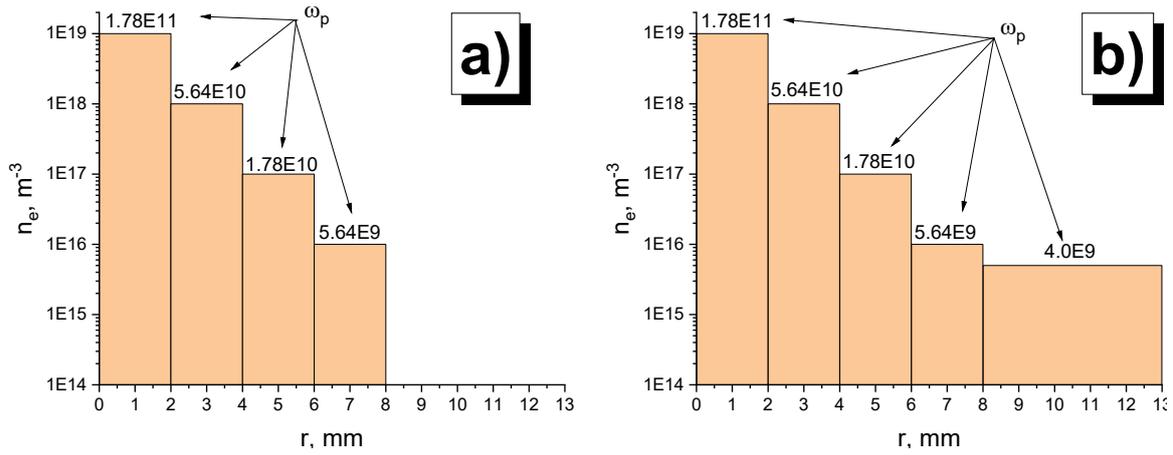

Figure 16. Electron density profiles for simulated geometry of a) Surfaguide-based experiment (see Fig. 14) and b) plasma torch at IPP (see Fig. 15). Plasma frequency ($\omega_p$) is labeled for corresponding radii and electron to neutral collision frequency is of $10^{11}$ s$^{-1}$.

The constructed $n_e(r)$ profiles for both experiments are shown in Fig. 16. The electron–neutral collision frequency was assumed to be constant over the radius, with a value of $1 \times 10^{11}$ s$^{-1}$. The input port in both simulated geometries was excited with microwave pulse signal of 200 ns in length and frequency of 2.45 GHz. The corresponding response signals from forward (blue) and back propagating (orange) waves at two ports of bi-directional coupler are shown in Fig. 17. The reflected signal is much less in amplitude compared to incident one as some power is absorbed in simulated plasma-object. The broadening of the reflected signal is evident and it is a known effect for pulse propagation in a waveguide [63]. Additionally, when the propagation path contains the elements, whose bandwidth is smaller than effective bandwidth of the incident microwave signal it initiates the reflections leading to the interference and, as a result, to amplitude perturbation in the beginning and in end of pulse. This modulation though needs to be distinguished from the observed much bigger relative variations in $P_{ref}(t)$ in the coaxial torch experiment, which originates from transient competition between load-mismatch reflection and electron density build-up after the breakdown.

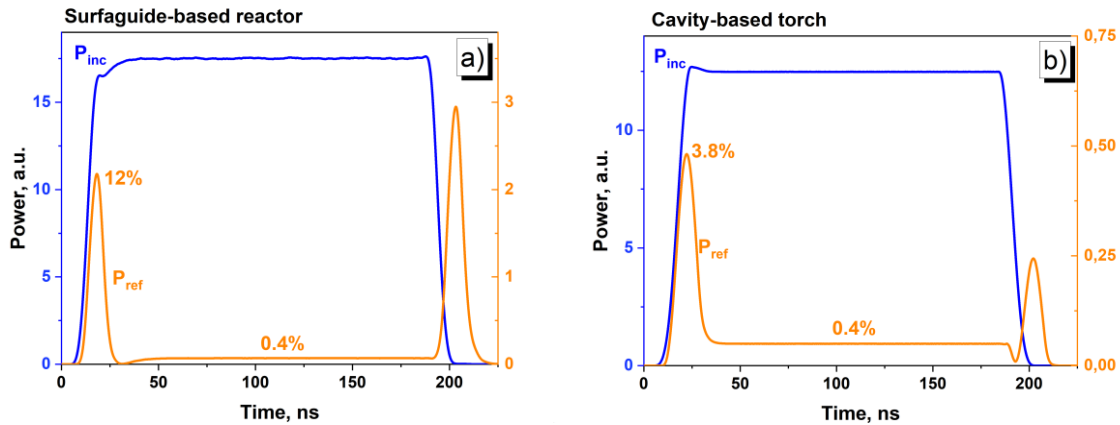

Figure 17. Incident and reflected power at outputs of bidirectional coupler simulated for a) Surfaguide-based experiment (see Fig. 14) and b) plasma torch at IPP (see Fig. 15) geometries. Frequency is 2.45 GHz, $t_{on}$ =200 ns and rising and falling fronts are 15 ns.

## 4. Summary and Conclusions

Atmospheric-pressure $CO_2$ plasma sustained by ultrafast kilowatt-range microwave pulsations was investigated in the Surfaguide-based reactor and cavity-based plasma torch at IPP and compared with pulsed plasmas generated in a compact coaxial torch. The Surfaguide-based and cavity-based plasma torch experiments employ comparable microwave power levels, coupling schemes, quartz reactor tubes, and gas injection geometries. The principal distinction between these systems lies in the post-discharge treatment: the plasma torch at IPP provides rapid, nozzle-driven quenching, whereas the Surfaguide-based reactor allows natural cooling along an extended quartz tube. This enables a direct assessment of the influence of afterglow dynamics on plasma–chemical performance. At the same time, both configurations differ markedly from the coaxial torch geometry, leading to distinct discharge regimes and energy transfer pathways. The design features, operating parameters, and performance characteristics of all three systems are summarized in table 4. Owing to the substantially lower *SEI* in the coaxial torch compared with the Surfaguide and IPP systems, absolute conversion rates differ significantly; therefore, the analysis focuses on relative performance changes under pulsed excitation compared with CW operation. Reignition phenomenon in pulsed scenarios in coaxial torch is considered as one of decisive factors which pushes its performance by more than 100% over the performance in CW scenario. Both the recorded $T_{vib}$-$T_{rot}$ nonequilibrium and increased plasma volume in pulsed mode evidence that plasma is more close to diffuse state, or less contracted as compared with plasmas in Surfaguide-based reactor and plasma torch at IPP. Despite the *SEI* operation range was too low and too far away from optimal 1-3 eV (due to restrictions for maximal power and minimal flow), the absolute efficiency of the process reached in coaxial torch of <27% is comparable with Surfaguide-based (<30%) and plasma torch at IPP (<27%) experiments. Note that the reignition regime was not attained, neither in Surfaguide-based nor in plasma torch at IPP experiments. In the Surfaguide-based reactor, within investigated duty cycles (*DC*>0.25) plasma was continuously burning both in pulse and inter-pulse phases, that was confirmed with time resolved $T_{vib}$, $T_{rot}$ and $P_{ref}$ diagnostics. Based on the analysis of instant reflected power, $P_{ref}$ we can suppose that also in cavity-based plasma torch at IPP the plasma has not extinguished during inter-pulse periods for considered 0.22<*DC*<0.26. Two limits were experimentally found for sustaining pulsed plasma in the Surfaguide-based reactor for a given peak power of about 4 kW: mean absorbed power $\overline{P_{abs}}$< 700 W and $t_{off}$ time > 12 μs. In both cases plasma gets extinguished. Similarly, in plasma torch at IPP the $t_{off}$ time > 10.3 μs was not accessible due to plasma instability. At the same time, the progress in Surfaguide-based reactor performance shown in Figure 8 looks quite promising, therefore we plan to further investigate plasma regimes with longer $t_{off}$ in future studies.

In present study, ultra-fast pulsations of MW power in the Surfaguide-based atmospheric pressure plasmas allow to increase plasma gas temperature, which appears to promote more advanced thermal dissociation: relative increase by 40 % and 20 % in conversion and efficiency, respectively, is documented. At the same time, the above progress due to enhanced $T_{gas}$ in pulsation mode was very likely possible because of relatively slow post-reactor quenching in the long quartz tube of the Surfaguide-based reactor, that was not the case in the cavity-based torch where gas temperature quenching in water cooled nozzle is rapid and massive. Apparently, the gas cooling trajectory and the gas cooling timing play an important role when thermally driven chemistry still take place in afterglow [57].

At the same time, we cannot exclude that fast energy pulsations may influence radial convection and diffusion fluxes of plasma species – specifically, the diffusion of CO molecules from the hot central plasma region toward the wall and of $CO_2$ molecules from the colder periphery toward the plasma center [64-66]. In modelling work [65] performed for a geometry similar to the two experiments considered here, turbulent fluxes were shown to be comparable to, or even exceed, molecular fluxes. Moreover, the maximum of turbulent kinetic energy, and thus the strongest influence on radial transport, was found to be shifted downstream from the heat source in the direction of the mean gas flow. For instance, the maximum turbulent thermal conductivity was predicted to occur approximately 100 mm downstream of the end of the heat source. In this regard, the Surfaguide-based reactor, which features a longer quenching trajectory compared with the IPP torch, could potentially benefit from enhanced turbulent transport induced by microwave pulsations. Yet, to what extent ultra-fast, microsecond pulsations can influence the radial transport balance is a topic for future investigations.

The analysis of the reflected microwave power provides at least qualitative insight into the dynamics of electron density in pulsed plasma. When the plasma is reignited with each incoming pulse, as in the coaxial torch, the growing electron population contributes to microwave power absorption and largely governs the reflected power signal, particularly at the beginning of the pulse. This agrees well with the results of Van Alphen *et al* (see Fig. S13-S16 in [34]) and van de Steeg *et al* [24], though obtained at vacuum conditions. The characteristic time for electron density development in the reignition regime is on the order of several tens of nanoseconds and also depends on the $t_{off}$ duration: the shorter the $t_{off}$, the faster the recovery of $n_e$. In contrast, in the Surfaguide-based reactor and the torch at IPP, the absence of strong modulation of $P_{ref}$ at the beginning of the pulse, together with the overall low mean level of reflected power, indicates that plasma reignition does not occur under these conditions. In future, the quantitative $n_e$ analysis would be also possible by accounting the complex reflection coefficient and varying the $n_e$ profile how it was realized in [67].

The microwave pulsation is believed to change the electron component in atmospheric $CO_2$ plasma. Depending on the reactor geometry, gas flow pattern, MW coupling and afterglow configuration it can be promising tool either for enhancing thermal chemistry or enabling transient $T_{vib}/T_{rot}$ nonequilibrium which helps to overcome (though partly) plasma contraction and promotes $CO_2$ conversion efficiency.

## Acknowledgements

The authors like to acknowledge Ante Hecimovic for organization and help in conduction of experiments at IPP site and for helpful discussions on the interpretation of observed physical effects. Vladislav Kotov is kindly acknowledged for insightful discussions on possible theoretical explanation of transient effects observed in pulsed plasma.

| | SEI, eV/molec | E/n, Td | Peak power & plasma volume | Afterglow & Quenching | Pulse timing | $n_e$ variation & Reignition | Effects from microwave pulsation ||||
|---|---|---|---|---|---|---|---|---|---|---|
| | | | | | | | Plasma temperatures / nonequilibrium | Relative enhancement as compared with CW | Absolute $\chi$ and $\eta$ in pulsed mode ||
| Coaxial torch | 0.02 – 0.22 | 65-75 | 220-260 W, $V_{plas}$=0.05-0.15 cm$^3$ | Gas expands in a bigger diameter tube | $t_{on}$= 0.5-50 µs DC= 0.10-0.99 | Strong $n_e$ variation due to reignition | $T_{rot}$= 3500-7500 K; $T_{vib}$= 6000-7500 K Nonequilibrium $T_{vib}/T_{rot} \approx 2$ [9] | $\Delta\chi$~100 %, $\Delta\eta$~130% $\Delta V_{plas}$~50-100% [9] | $\chi$~1 %, $\eta$~27% |
| Surfaguide-based reactor (KIT) | 0.7 – 2.7 | 10-40 | 3.0-4.2 kW $V_{plas}$<30 cm$^3$ including expansion above the reactor | Slow moderate cooling in long quartz tube | $t_{on}$= 1.0-10 µs DC= 0.25-0.9 | Seemingly small $n_e$ variation based on $P_{ref}(t)$ No reignition | $T_{rot}$= 6000-7000 K $T_{vib}$ = 7000-8500 K | $\Delta\chi$~40% $\Delta\eta$~20% $\Delta T_{gas}$ ~13% | $\chi$~15 % $\eta$~25 % |
| Cavity-based plasma torch (IPP) | 1.5 – 5.0 | 10-40 | 4.05 kW $V_{plas} \approx 23$ cm$^3$ | Fast, quick quenching in water-cooled nozzle | $t_{on}$= 0.5-10 µs DC= 0.22-0.62 | Seemingly small $n_e$ variation based on $P_{ref}(t)$ No reignition | Not measured | No progress in $\chi$ and $\eta$ | $\chi$~20 % $\eta$~25 % |

Table 4. Comparison of setups, main parameters and performance of all three experiments considered.